\documentclass[journal,twoside]{IEEEtran}

\usepackage{cite}
\usepackage{graphicx}
\usepackage{balance}
\usepackage{setspace}

\begin{document}

\title{Finite-Difference Time-Domain Study of Guided Modes in Nano-plasmonic Waveguides}

\author{Yan~Zhao,~\IEEEmembership{Student~Member,~IEEE,}
        and~Yang~Hao,~\IEEEmembership{Senior~Member,~IEEE}}

\maketitle

\begin{abstract}
A conformal dispersive finite-difference time-domain (FDTD) method
is developed for the study of one-dimensional (1-D) plasmonic
waveguides formed by an array of periodic infinite-long silver
cylinders at optical frequencies. The curved surfaces of circular
and elliptical inclusions are modelled in orthogonal FDTD grid using
effective permittivities (EPs) and the material frequency dispersion
is taken into account using an auxiliary differential equation (ADE)
method. The proposed FDTD method does not introduce numerical
instability but it requires a fourth-order discretisation procedure.
To the authors' knowledge, it is the first time that the modelling
of curved structures using a conformal scheme is combined with the
dispersive FDTD method. The dispersion diagrams obtained using EPs
and staircase approximations are compared with those from the
frequency domain embedding method. It is shown that the dispersion
diagram can be modified by adding additional elements or changing
geometry of inclusions. Numerical simulations of plasmonic
waveguides formed by seven elements show that row(s) of silver
nanoscale cylinders can guide the propagation of light due to the
coupling of surface plasmons.
\end{abstract}

\section{Introduction}
It is well known that photonic crystals (PCs) offer unique
opportunities to control the flow of light \cite{Joannopoulos}. The
basic idea is to design periodic dielectric structures that have a
bandgap for a particular frequency range. Periodic dielectric rods
with removed one or several rows of elements can be used as
waveguiding devices when operating at bandgap frequencies. A lot of
effort has been made to obtain a complete and wider bandgap. It has
been shown that a triangular lattice of air holes in a dielectric
background has a complete bandgap for TE (transverse electric) mode,
while a square lattice of dielectric rods in air has a bandgap for
TM (transverse magnetic) mode \cite{Qiu1}. The devices operating in
the bandgap frequencies are not the only option to guide the flow of
light. Another waveguiding mechanism is the total internal
reflection (TIR) in one-dimensional (1-D) periodic dielectric rods
\cite{Fan1}. It is analysed in \cite{Fan1} that a single row of
dielectric rods or air holes supports waveguiding modes and
therefore can be also used as waveguide. In \cite{Chigrin}, the
design of such waveguides consisting of several rows of dielectric
rods with various spacings is proposed.

Recently, a new method for guiding electromagnetic waves in
structures whose dimensions are below the diffraction limit has been
proposed. The structures are termed as `plasmonic waveguides' which
have an operation of principle based on near-field interactions
between closely spaced noble metal nanoparticles (spacing
$\ll\lambda$) that can be efficiently excited at their surface
plasmon frequency. The guiding principle relies on coupled plasmon
modes set up by near-field dipole interactions that lead to coherent
propagation of energy along the array. Analogous structures as
waveguides in microwave regime include periodic metallic cylinders
to support propagating waves \cite{Shefer}, array of flat dipoles
which support guided waves \cite{Munk}, and Yagi-Uda antennas
\cite{Mailloux,Yaghjian} etc. However, although these structures can
be scaled to optical frequencies with appropriate material
properties, their dimensions are limited by the so-called
diffraction limit $\lambda/(2n)$. On the other hand, plasmonic
waveguides employ the localisation of electromagnetic fields near
metal surfaces to confine and guide light in regions much smaller
than the free space wavelength and can effectively overcome the
diffraction limit. Previous analysis of plasmonic structures
includes the plasmon propagation along metal stripes, wires, or
grooves in metal
\cite{Weeber,Lamprecht,Yatsui,Zia,Charbonneau,Pile}, and the
coupling between plasmons on metal particles in order to guide
energy \cite{Quinten,Brongersma} etc. Such subwavelength structures
can also find their applications e.g. efficient absorbers and
electrically small receiving antennas at microwave frequencies.
Recently composite materials containing randomly distributed
electrically conductive material and non-electrically conductive
material have been designed \cite{Youngs}. They are noted to exhibit
plasma-like responses at frequencies well below plasma frequencies
of the bulk material.

The finite-difference time-domain (FDTD) method \cite{Taflove} is
seen as the most popular numerical technique especially because of
its flexibility in handling material dispersion as well as arbitrary
shaped inclusions. In \cite{Maier}, the optical pulse propagation
below the diffraction limit is shown using the FDTD method. Also
with the FDTD method, the waveguide formed by several rows of silver
nanorods arranged in hexagonal is studied \cite{Saj}. Despite these
examples of applying the FDTD method for the plasmonic structures,
the accuracy of modelling has not been proven yet. When modelling
curved structures, unless using extremely fine meshes, due to the
nature of orthogonal and staggered grid of conventional FDTD, often
modifications need to be applied in order to improve the numerical
accuracy, such as the treatment of interfaces between different
materials even for planar structures \cite{Hwang}, and the improved
conformal algorithms using structured meshes \cite{HaoConformal} for
curved surfaces.

In addition to the modifications at material interfaces, the
material frequency dispersion has also to be taken into account in
FDTD modelling \cite{Luebbers1,Gandhi1,Sullivan1}. However,
modelling dispersive materials with curved surfaces still remains to
be a challenging topic due to the complexity in algorithm as well as
the introduction of numerical instability. An alternative way to
solve this problem is based on the idea of effective permittivities
(EPs) \cite{Kaneda,Lee,Mohammadi} in the underlying Cartesian
coordinate system, and the dispersive FDTD scheme can be therefore
modified accordingly without affecting the stability of algorithm.
In this paper, we first propose a novel conformal dispersive FDTD
algorithm combining the EPs together with an auxiliary differential
equation (ADE) method \cite{Taflove}, then apply the developed
method to the modelling of plasmonic waveguides formed by an array
of circular or elliptical shaped silver cylinders at optical
frequencies. Numerical FDTD simulation results are verified by a
frequency domain embedding method \cite{Inglesfield}. To the
authors' knowledge, it is the first time that a conformal scheme is
combined with the dispersive FDTD method for the modelling of
nano-plasmonic waveguides.

\section{Conformal Dispersive FDTD Method Using Effective Permittivities}
\begin{figure}[t]
\centering
\includegraphics[width=8.6cm]{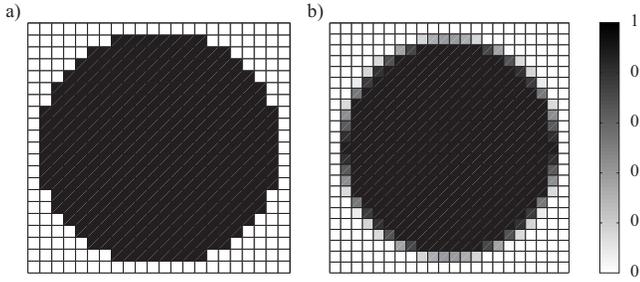}
\caption{Comparison of the filling ratio for $E_y$ component in FDTD
modelling of a circular cylinder using (a) staircase approximations
and (b) a conformal scheme. The radius of circular cylinder is ten
cells.} \label{fig_filling_ratio}
\end{figure}

Conventionally, staircase approximations are often used to model
curved electromagnetic structures in an orthogonal FDTD domain.
Figure~\ref{fig_filling_ratio}(a) shows an example layout of an
infinite-long cylinder in the free space represented in a
two-dimensional (2-D) orthogonal FDTD domain.
The approximated shape introduces spurious numerical resonant modes
which do not exist in actual structures. On the other hand, using
the concept of filling ratio, which is defined as the ratio of the
area of material $\varepsilon_2$ to the area of a particular FDTD
cell, the curvature can be properly represented in FDTD domain as
shown in Fig.~\ref{fig_filling_ratio}(b), where different levels of
darkness indicate different filling ratios of material
$\varepsilon_2$. The accuracy of modelling can be significantly
improved compared with staircase approximations, as will be shown in
a later section.

\begin{figure}[t]
\centering
\includegraphics[width=6.5cm]{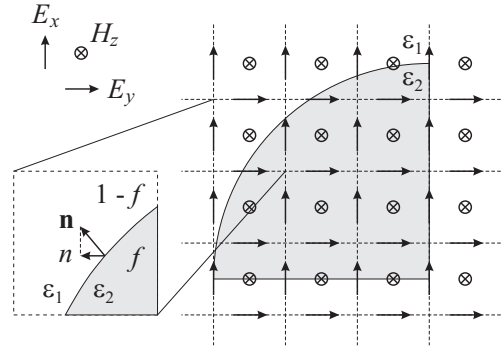}
\caption{Layout of a quarter circular inclusion in orthogonal FDTD
grid for $E_y$ component. The radius of circular cylinder is three
cells.} \label{fig_grid}
\end{figure}

According to \cite{Mohammadi}, the EP in a general form is given by
\begin{equation}
\varepsilon_{\rm{eff}}=\varepsilon_\parallel(1-n^2)+\varepsilon_\perp
n^2, \label{eq_per_eff}
\end{equation}
where $n$ is the projection of the unit normal vector $\textbf{n}$
along the field component as shown in Fig.~\ref{fig_grid},
$\varepsilon_\parallel$ and $\varepsilon_\perp$ are parallel and
perpendicular permittivities to the material interface, respectively
and defined as
\begin{eqnarray}
\varepsilon_\parallel&=&f\varepsilon_2+(1-f)\varepsilon_1, \label{eq_per_s}\\
\varepsilon_\perp&=&\left[f/\varepsilon_2+(1-f)/\varepsilon_1\right]^{-1},
\label{eq_per_p}
\end{eqnarray}
where $f$ is the filling ratio of material $\varepsilon_2$ in a
certain FDTD cell.

In this paper we consider the inclusions as silver cylinders, which
at optical frequencies can be modelled using the Drude dispersion
model
\begin{equation}
\varepsilon_2(\omega)=\varepsilon_0\left(1-\frac{\omega^2_p}{\omega^2-j\omega\gamma}\right),
\label{eq_per_2}
\end{equation}
where $\omega_p$ is the plasma frequency and $\gamma$ is the
collision frequency. At the frequencies below the plasma frequency,
the real part of permittivity is negative. In this paper, we assume
that the silver cylinders are embedded in the free space
($\varepsilon_1=\varepsilon_0$).

In order to take into account the frequency dispersion of the
material, the electric flux density \textbf{D} is introduced into
standard FDTD updating equations. At each time step, \textbf{D} is
updated directly from \textbf{H} and \textbf{E} can be calculated
from \textbf{D} through the following steps. Substitute
(\ref{eq_per_s}) and (\ref{eq_per_p}) into (\ref{eq_per_eff}) and
using the expressions for $\varepsilon_1$ and $\varepsilon_2$
(\ref{eq_per_2}), the constitutive relation in the frequency domain
reads
\begin{eqnarray}
&&\!\!\!\!\!\!\!\!\!\!\!\!\{\omega^4-2\gamma
j\omega^3-[\gamma^2+(1-f)\omega^2_p]\omega^2+\gamma(1-f)\omega^2_p
j\omega\}\textbf{D}\nonumber\\
&&\!\!\!\!\!\!\!\!\!\!\!\!=[\omega^4-2\gamma
j\omega^3-(\gamma^2+\omega^2_p)\omega^2+\gamma\omega^2_p
j\omega\nonumber\\
&&\!\!\!\!+f(1-f)(1-n^2)\omega^4_p]\varepsilon_0\textbf{E}.
\label{eq_constitutive}
\end{eqnarray}
Using the inverse Fourier transformation i.e.
$j\omega\rightarrow\partial/\partial t$, we obtain the constitutive
relation in the time domain as
\begin{eqnarray}
&&\!\!\!\!\!\!\!\!\!\!\!\!\left\{\frac{\partial^4}{\partial
t^4}+2\gamma\frac{\partial^3}{\partial
t^3}+[\gamma^2+(1-f)\omega^2_p]\frac{\partial^2}{\partial
t^2}+\gamma(1-f)\omega^2_p\frac{\partial}{\partial t}\right\}\textbf{D}\nonumber\\
&&\!\!\!\!\!\!\!\!\!\!\!\!=\bigg[\frac{\partial^4}{\partial
t^4}+2\gamma\frac{\partial^3}{\partial
t^3}+(\gamma^2+\omega^2_p)\frac{\partial^2}{\partial
t^2}+\gamma\omega^2_p\frac{\partial}{\partial t}\nonumber\\
&&\!\!\!\!+f(1-f)(1-n^2)\omega^4_p\bigg]\varepsilon_0\textbf{E}.
\label{eq_differential}
\end{eqnarray}

The FDTD simulation domain is represented by an equally spaced
three-dimensional (3-D) grid with periods $\Delta x$, $\Delta y$ and
$\Delta z$ along $x$-, $y$- and $z$-directions, respectively. The
time step is $\Delta t$. For discretisation of
(\ref{eq_differential}), we use the central finite difference
operators in time ($\delta_t$) and the central average operator with
respect to time ($\mu_t$):
\begin{eqnarray}
&&\!\!\!\!\!\!\!\!\!\frac{\partial^4}{\partial
t^4}\rightarrow\frac{\delta^4_t}{(\Delta
t)^4},~~~\frac{\partial^3}{\partial
t^3}\rightarrow\frac{\delta^3_t}{(\Delta
t)^3}\mu_t,~~~\frac{\partial^2}{\partial
t^2}\rightarrow\frac{\delta^2_t}{(\Delta t)^2}\mu^2_t,\nonumber\\
&&\!\!\!\!\!\!\!\!\!\frac{\partial}{\partial
t}\rightarrow\frac{\delta_t}{\Delta
t}\mu^3_t,~~~1\rightarrow\mu^4_t, \label{eq_operator}
\end{eqnarray}
where the operators $\delta_t$ and $\mu_t$ are defined as in
\cite{Hildebrand}:
\begin{eqnarray}
\delta_t\textbf{F}|^n_{m_x,m_y,m_z}&\equiv&\textbf{F}|^{n+\frac{1}{2}}_{m_x,m_y,m_z}-\textbf{F}|^{n-\frac{1}{2}}_{m_x,m_y,m_z},\\
\mu_t\textbf{F}|^n_{m_x,m_y,m_z}&\equiv&\frac{\textbf{F}|^{n+\frac{1}{2}}_{m_x,m_y,m_z}+\textbf{F}|^{n-\frac{1}{2}}_{m_x,m_y,m_z}}{2}.
\label{eq_operators}
\end{eqnarray}
Here $\textbf{F}$ represents field components and
$m_{x},m_{y},m_{z}$ are indices corresponding to a certain
discretisation point in the FDTD domain. The discretised Eq.
(\ref{eq_differential}) reads
\begin{eqnarray}
&&\!\!\!\!\!\!\!\!\!\!\!\!\bigg\{\frac{\delta^4_t}{(\Delta
t)^4}+2\gamma\frac{\delta^3_t}{(\Delta
t)^3}\mu_t+\left[\gamma^2+(1-f)\omega^2_p\right]\frac{\delta^2_t}{(\Delta
t)^2}\mu^2_t\nonumber\\
&&\!\!\!\!\!\!\!+\gamma(1-f)\omega^2_p\frac{\delta_t}{\Delta
t}\mu^3_t\bigg\}\textbf{D}=\bigg[\frac{\delta^4_t}{(\Delta
t)^4}+2\gamma\frac{\delta^3_t}{(\Delta
t)^3}\mu_t\nonumber\\
&&\!\!\!\!\!\!\!+(\gamma^2+\omega^2_p)\frac{\delta^2_t}{(\Delta
t)^2}\mu^2_t+\gamma\omega^2_p\frac{\delta_t}{\Delta
t}\mu^3_t\nonumber\\
&&\!\!\!\!\!\!\!+f(1-f)(1-n^2)\omega^4_p\mu^4_t\bigg]\varepsilon_0\textbf{E}.
\label{eq_differential_approx}
\end{eqnarray}
Note that in the above equations we have kept all terms to be the
fourth-order to guarantee numerical stability. Equation
(\ref{eq_differential_approx}) can be written as
\begin{eqnarray}
&&\!\!\!\frac{\textbf{D}^{n+1}-4\textbf{D}^n+6\textbf{D}^{n-1}-4\textbf{D}^{n-2}
+\textbf{D}^{n-3}}{(\Delta t)^4}\nonumber\\
&&~+\gamma\frac{\textbf{D}^{n+1}-2\textbf{D}^n+2\textbf{D}^{n-2}-\textbf{D}^{n-3}}{(\Delta t)^3}\nonumber\\
&&~+\left[\gamma^2+(1-f)\omega^2_p\right]\frac{\textbf{D}^{n+1}-2\textbf{D}^{n-1}+\textbf{D}^{n-3}}{4(\Delta
t)^2}\nonumber\\
&&~+\gamma(1-f)\omega^2_p\frac{\textbf{D}^{n+1}+2\textbf{D}^n-2\textbf{D}^{n-2}-\textbf{D}^{n-3}}{8\Delta
t}\nonumber\\
&&\!\!\!=\varepsilon_0\frac{\textbf{E}^{n+1}-4\textbf{E}^n+6\textbf{E}^{n-1}-4\textbf{E}^{n-2}+\textbf{E}^{n-3}}{(\Delta
t)^4}\nonumber\\
&&~+\varepsilon_0\gamma\frac{\textbf{E}^{n+1}-2\textbf{E}^n+2\textbf{E}^{n-2}-\textbf{E}^{n-3}}{(\Delta t)^3}\nonumber\\
&&~+\varepsilon_0\left(\gamma^2+\omega^2_p\right)\frac{\textbf{E}^{n+1}-2\textbf{E}^{n-1}+\textbf{E}^{n-3}}{4(\Delta
t)^2}\nonumber\\
&&~+\varepsilon_0\gamma\omega^2_p\frac{\textbf{E}^{n+1}+2\textbf{E}^n-2\textbf{E}^{n-2}-\textbf{E}^{n-3}}{8\Delta
t}\nonumber\\
&&~+\frac{\varepsilon_0f(1-f)(1-n^2)\omega^4_p}{16}\big(\textbf{E}^{n+1}+4\textbf{E}^n+6\textbf{E}^{n-1}\nonumber\\
&&~~~~+4\textbf{E}^{n-2}+\textbf{E}^{n-3}\big).
\label{eq_constitutive_approx2}
\end{eqnarray}
The indices $m_x$, $m_y$ and $m_z$ are omitted from
(\ref{eq_constitutive_approx2}) since \textbf{E} and \textbf{D} are
located at same locations. Solving for $\textbf{E}^{n+1}$ then the
updating equation for $\textbf{E}$ in FDTD iterations reads
\begin{eqnarray}
&&\!\!\!\!\!\!\!\!\!\!\!\!\!\!\!\!\!\!\!\textbf{E}^{n+1}=[b_0\textbf{D}^{n+1}+b_1\textbf{D}^n+b_2\textbf{D}^{n-1}+b_3\textbf{D}^{n-2}+b_4\textbf{D}^{n-3}\nonumber\\
&&~-(a_1\textbf{E}^n+a_2\textbf{E}^{n-1}+a_3\textbf{E}^{n-2}+a_4\textbf{E}^{n-3})]/a_0,
\label{eq_update}
\end{eqnarray}
with the coefficients given by
\begin{eqnarray}
a_0\!\!\!\!&=&\!\!\!\!\varepsilon_0\Bigg[\frac{1}{(\Delta
t)^4}+\frac{\gamma}{(\Delta
t)^3}+\frac{\gamma^2+\omega^2_p}{4(\Delta
t)^2}+\frac{\gamma\omega^2_p}{8\Delta t}\nonumber\\
&&~~~~+\frac{f(1-f)(1-n^2)\omega^4_p}{16}\Bigg],\nonumber\\
a_1\!\!\!\!&=&\!\!\!\!\varepsilon_0\left[-\frac{4}{(\Delta
t)^4}-\frac{2\gamma}{(\Delta t)^3}+\frac{\gamma\omega^2_p}{4\Delta
t}+\frac{f(1-f)(1-n^2)\omega^4_p}{4}\right],\nonumber\\
a_2\!\!\!\!&=&\!\!\!\!\varepsilon_0\left[\frac{6}{(\Delta
t)^4}-\frac{\gamma^2+\omega^2_p}{2(\Delta t)^2}+\frac{3f(1-f)(1-n^2)\omega^4_p}{8}\right],\nonumber\\
a_3\!\!\!\!&=&\!\!\!\!\varepsilon_0\left[-\frac{4}{(\Delta
t)^4}+\frac{2\gamma}{(\Delta
t)^3}-\frac{\gamma\omega^2_p}{4\Delta t}+\frac{f(1-f)(1-n^2)\omega^4_p}{4}\right],\nonumber\\
a_4\!\!\!\!&=&\!\!\!\!\varepsilon_0\Bigg[\frac{1}{(\Delta
t)^4}-\frac{\gamma}{(\Delta
t)^3}+\frac{\gamma^2+\omega^2_p}{4(\Delta
t)^2}-\frac{\gamma\omega^2_p}{8\Delta
t}\nonumber\\
&&~~~~+\frac{f(1-f)(1-n^2)\omega^4_p}{16}\Bigg],\nonumber\\
b_0\!\!\!\!&=&\!\!\!\!\frac{1}{(\Delta t)^4}+\frac{\gamma}{(\Delta
t)^3}+\frac{\gamma^2+(1-f)\omega^2_p}{4(\Delta
t)^2}+\frac{\gamma(1-f)\omega^2_p}{8\Delta
t},\nonumber\\
b_1\!\!\!\!&=&\!\!\!\!-\frac{4}{(\Delta t)^4}-\frac{2\gamma}{(\Delta
t)^3}+\frac{\gamma(1-f)\omega^2_p}{4\Delta
t},\nonumber\\
b_2\!\!\!\!&=&\!\!\!\!\frac{6}{(\Delta
t)^4}-\frac{\gamma^2+(1-f)\omega^2_p}{2(\Delta
t)^2},\nonumber\\
b_3\!\!\!\!&=&\!\!\!\!-\frac{4}{(\Delta t)^4}+\frac{2\gamma}{(\Delta
t)^3}-\frac{\gamma(1-f)\omega^2_p}{4\Delta
t},\nonumber\\
b_4\!\!\!\!&=&\!\!\!\!\frac{1}{(\Delta t)^4}-\frac{\gamma}{(\Delta
t)^3}+\frac{\gamma^2+(1-f)\omega^2_p}{4(\Delta
t)^2}-\frac{\gamma(1-f)\omega^2_p}{8\Delta t},\nonumber
\label{eq_coefficients}
\end{eqnarray}
The computations of $\textbf{H}$ and $\textbf{D}$ are performed
using Yee's standard updating equations in the free space. Note that
if the plasma frequency is equal to zero ($\omega_p=0$), then
(\ref{eq_update}) reduces to the updating equation in the free space
i.e. $\textbf{E}=\textbf{D}/\varepsilon_0$.

\section{FDTD Calculation of Dispersion Diagram}
Applying the Bloch's periodic boundary conditions (PBCs)
\cite{ChanPBC,Holter,Turner,Prescott,Ren,Roden}, FDTD method can be
used to model periodic structures and calculate their dispersion
diagrams \cite{Fan2,Qiu2}. For any periodic structures, the field at
any time should satisfy the Bloch theory, i.e.
\begin{equation}
\textbf{E}(\textbf{d}+\textbf{a})=\textbf{E}(\textbf{d})e^{j\bf{ka}},~~~\textbf{H}(\textbf{d}+\textbf{a})=\textbf{H}(\textbf{d})e^{j\bf{ka}},
\label{eq_PBC}
\end{equation}
where $\textbf{d}$ is the distance vector of any location in the
computation domain, $\textbf{k}$ is the wave vector and $\textbf{a}$
is the lattice vector along the direction of periodicity. When
updating the fields at the boundary of the computation domain using
FDTD, the required fields outside the computation domain can be
calculated using known field values inside the domain through Eq.
(\ref{eq_PBC}). Although instead of using real values in
conventional FDTD computations, the calculation of dispersion
diagrams requires complex field values, since only one unit cell is
modelled, the computation load is not significantly increased.

\begin{figure}[t]
\centering \includegraphics[width=7cm]{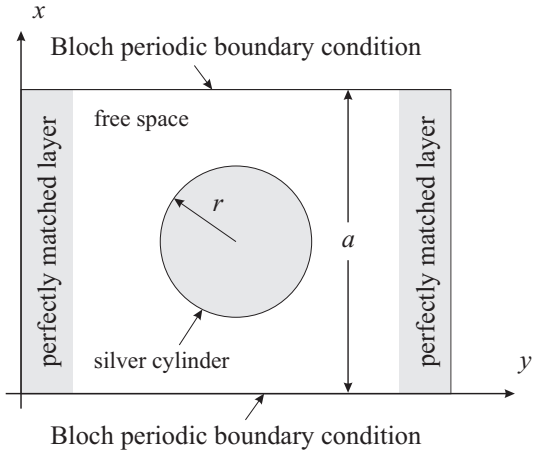} \caption{The
layout of the 2-D FDTD computation domain for calculating dispersion
diagram for 1-D periodic structures. The inclusion has a circular
cross-section with radius $r$ and the period of the 1-D infinite
structure is $a$.} \label{fig_domain}
\end{figure}

First we apply the developed conformal dispersive FDTD method to
calculate the dispersion diagram for 1-D plasmonic waveguides formed
by an array of periodic infinite-long (along $z$-direction) circular
silver cylinders. As shown in Fig.~\ref{fig_domain}, the 2-D
simulation domain ($x$-$y$) with TE modes (therefore only $E_x$,
$E_y$ and $H_z$ are non-zero fields) is truncated using Bloch's PBCs
in $x$-direction and Berenger's perfectly matched layers (PMLs)
\cite{Berenger} in $y$-direction.
The Berenger's PMLs have excellent performance for absorbing
propagating waves \cite{Berenger}, however for evanescent waves,
field shows growing behaviour inside PMLs. Since the waves radiated
by point or line sources consist of both propagating and evanescent
components, extra space (typically a quarter wavelength at the
frequency of interest) between PMLs and the circular inclusion is
added to allow the evanescent waves to decay before reaching the
PMLs.

The radius of silver cylinders is $r=2.5\times10^{-8}$ m and the
period is $a=7.5\times10^{-8}$ m. The plasma and collision
frequencies are $\omega_p=9.39\times10^{15}$ rad/s and
$\gamma=3.14\times10^{13}$ Hz, respectively in order to closely
match the bulk dielectric function of silver \cite{Johnson}. The
FDTD cell size is $\Delta x=\Delta y=2.5\times10^{-9}$ m with the
time step $\Delta t=\Delta x/(\sqrt{2}c)$ s (where $c$ is the speed
of light in the free space) according to the Courant stability
criterion \cite{Taflove}. Although the stability condition for
high-order FDTD method is typically more strict than the
conventional one, since the average operator $\mu_t$ is applied to
develop the algorithm, we have not found any instability for a
complete time period of more than 40,000 time steps used in all
simulations.

A wideband magnetic line source is placed at an arbitrary location
in the free space region of the 2-D simulation domain in order to
excite all resonant modes of the structure within the frequency
range of interest (normalised frequency $\overline f=\omega a/(2\pi
c)\in[0\sim0.5]$):
\begin{equation}
g(t)=e^{-\left(\frac{t-t_{0}}{\tau}\right)^{2}}\cdot e^{j\omega t}
\label{eq_Gaussian}
\end{equation}
where $t_{0}$ is the initial time delay, $\tau$ defines the pulse
width and $\omega$ is the centre frequency of the pulse ($\overline
f=0.25$). The magnetic fields at one hundred random locations in the
free space region are recorded during simulations, transformed into
the frequency domain and combined to extract individual resonant
mode corresponding to each local maximum. For each wave vector, a
total number of 40,000 time steps are used in our simulations to
obtain enough accurate frequency domain results.

\begin{figure}[t]
\centering
\includegraphics[width=8.6cm]{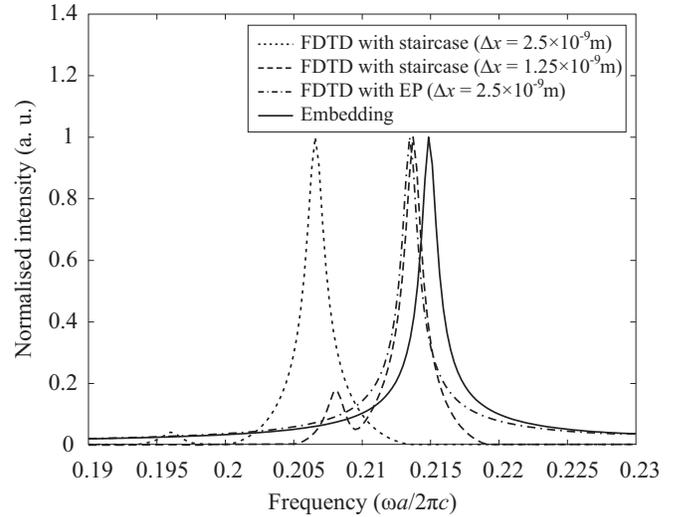}
\caption{Comparison of the first resonant frequency (transverse
mode) at wave vector $k_x=\pi/a$ calculated using the FDTD method
with staircase approximations, the FDTD method with EPs and the
frequency domain embedding method.} \label{fig_resonance}
\end{figure}

In order to demonstrate the advantage of EPs and validate the
proposed conformal dispersive FDTD method, we have also performed
simulations using staircase approximations for the circular
cylinder, as shown in Fig.~\ref{fig_filling_ratio}(a).
Figure~\ref{fig_resonance} shows the comparison of the first
resonant frequency (transverse mode) at wave vector $k_x=\pi/a$ of
the plasmonic waveguide calculated using the FDTD method with
staircase approximations, the FDTD method with EPs and the frequency
domain embedding method \cite{Giannakis}.
With the same FDTD spatial resolutions, the model using EP shows
excellent agreement with the results from the frequency domain
embedding method, on the contrary, the staircase approximation not
only leads to a shift of the main resonant frequency, but also
introduces a spurious numerical resonant mode which does not exist
in actual structures. The same effect has also been found for
non-dispersive dielectric cylinders \cite{Song}. It is also shown in
Fig.~\ref{fig_resonance} that although one may correct the main
resonant frequency using finer meshes, the spurious resonant mode
still remains.

\begin{figure}[t]
\centering
\includegraphics[width=8.6cm]{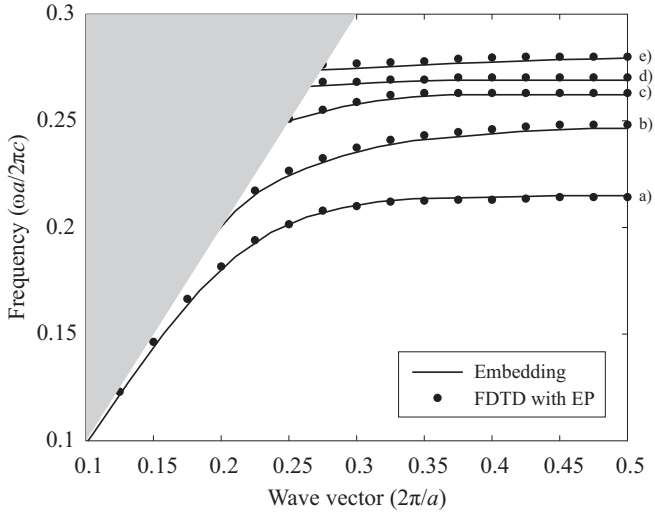}
\caption{Comparison of dispersion diagrams for an array of
infinite-long (along $z$-direction) circular silver cylinders
calculated using the FDTD method with EPs and the frequency domain
embedding method.} \label{fig_dispersion_1circular}
\end{figure}

The problem of frequency shift and spurious modes become severer
when calculating higher guided modes near the `flat band' region
(i.e the region where waves travel at a very low phase velocity).
Even with a refined spatial resolution, the staircase approximation
fails to provide correct results (not shown). On the other hand,
using the proposed conformal dispersive FDTD scheme, all resonant
modes are correctly captured in FDTD simulations as demonstrated by
the comparison with the embedding method as shown in
Fig.~\ref{fig_dispersion_1circular}.

\begin{figure}[t]
\centering
\includegraphics[width=8.6cm]{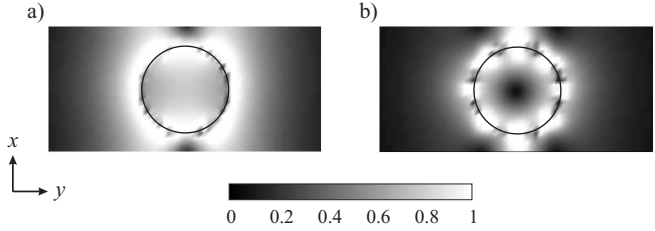}
\caption{Normalised total electric field intensities corresponding
to (a) transverse and (b) longitudinal modes \cite{Giannakis} at
wave number $k_x=\pi/a$ as marked in
Fig.~\ref{fig_dispersion_1circular}. The structure is infinite along
$x$-direction.} \label{fig_field_1circular_E}
\end{figure}

\begin{figure}[t]
\centering
\includegraphics[width=8.6cm]{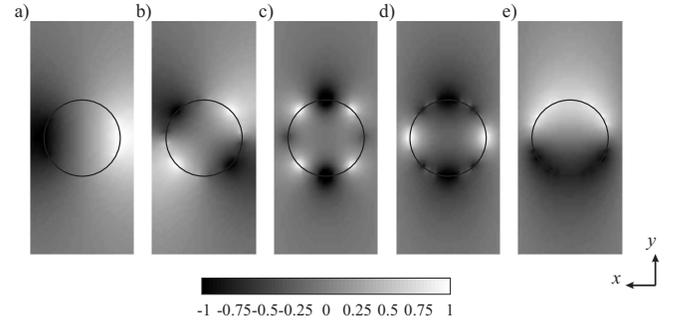}
\caption{Normalised distributions of magnetic field corresponding to
different resonant modes at wave number $k_x=\pi/a$ as marked in
Fig.~\ref{fig_dispersion_1circular}: (a), (c), (d): even modes, and
(b), (e): odd modes. The structure is infinite along $x$-direction.
(Note that the coordinate has been rotated 90 degrees anti-clockwise
from Fig.~\ref{fig_domain} for better presentation of the figure.)}
\label{fig_field_1circular}
\end{figure}

According to previous analysis using the frequency embedding method,
the fundamental mode in the modelled plasmonic waveguide is
transverse mode and the second guided mode is longitudinal
\cite{Giannakis}, which is also shown by the distribution of
electric field intensities in Fig.~\ref{fig_field_1circular_E} from
our FDTD simulations.
The higher guided modes are referred to as plasmon modes. For
demonstration of field symmetries and due to the TE mode considered
in our simulations, we have plotted the distributions of magnetic
field corresponding to different resonant modes at wave number
$k_x=\pi/a$ as marked in Fig.~\ref{fig_dispersion_1circular}, as
shown in Fig.~\ref{fig_field_1circular}.
Sinusoidal sources for excitation of certain single mode are used
and the sources are placed at different locations corresponding to
different symmetries of field patterns. All field patterns are
plotted after the steady state is reached in simulations. The modes
(a), (c) and (d) in Fig.~\ref{fig_field_1circular} are even modes
(relative to the direction of periodicity of the waveguide i.e.
$x$-axis), and (b) and (e) are considered as odd modes.

\begin{figure}[t]
\centering
\includegraphics[width=8.6cm]{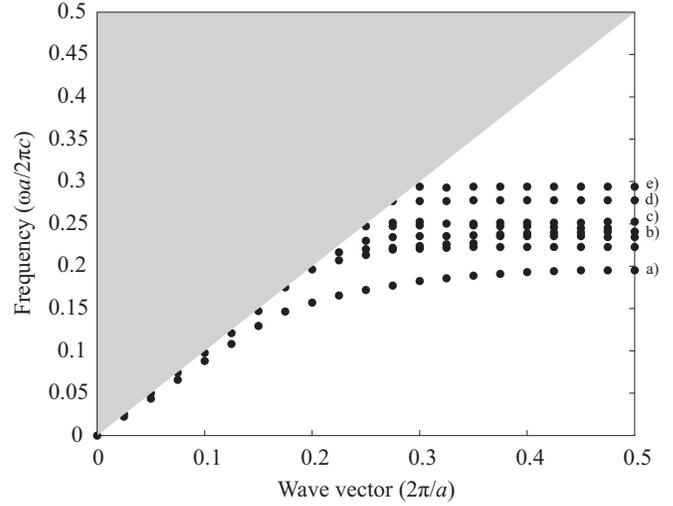}
\caption{Dispersion diagram for two rows of periodic infinite-long
(along $z$-direction) circular silver cylinders arranged in square
lattice calculated from conformal dispersive FDTD simulations.}
\label{fig_dispersion_2circular}
\end{figure}

\begin{figure}[t]
\centering
\includegraphics[width=8.6cm]{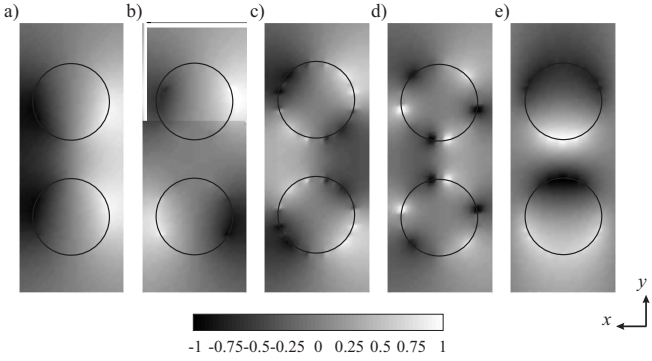}
\caption{Normalised distributions of magnetic field corresponding to
different guided modes as marked in
Fig.~\ref{fig_dispersion_2circular}: (a), (c), (d): even modes, and
(b), (e): odd modes. The structure is infinite along $x$-direction.
(Note that the coordinate has been rotated 90 degrees anti-clockwise
from Fig.~\ref{fig_domain} for better presentation of the figure.)}
\label{fig_field_2circular}
\end{figure}

\begin{figure}[t]
\centering
\includegraphics[width=8.6cm]{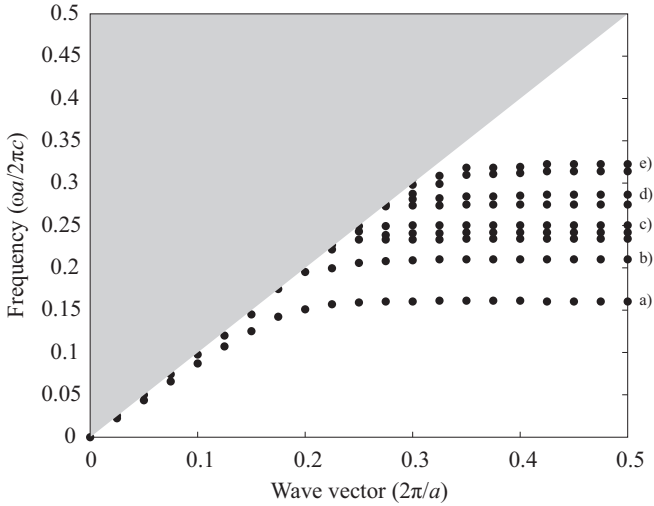}
\caption{Dispersion diagram for a single row of periodic
infinite-long (along $z$-direction) elliptical silver cylinders
calculated from conformal dispersive FDTD simulations.}
\label{fig_dispersion_elliptical}
\end{figure}

The above comparison of the simulation results calculated using the
conformal dispersive FDTD method and the embedding method clearly
demonstrates the effectiveness of applying the EPs in FDTD
modelling. Furthermore, in contrast to the embedding method, the
main advantage of the FDTD method is that arbitrary shaped
geometries can be easily modelled. We have applied the conformal
dispersive FDTD method to study the effect of different inclusions
on the dispersion diagrams of 1-D plasmonic waveguides. The
geometries considered are two rows of periodic infinite-long (along
$z$-direction) circular silver cylinders arranged in square lattice
and a single row of periodic infinite-long (along $z$-direction)
elliptical silver cylinders. The elliptical element has a ratio of
semimajor-to-semiminor axis 2:1, where the semiminor axis is equal
to the radius of the circular element ($25.0$ nm). For the two rows
of circular nanorods, the spacing between two rows (centre-to-centre
distance) is $75.0$ nm. The dispersion diagrams for these structures
are plotted in Figs.~\ref{fig_dispersion_2circular} and
\ref{fig_dispersion_elliptical}.
Comparing the dispersion diagrams for a single circular element in
Fig.~\ref{fig_dispersion_1circular} and two circular elements in
Fig.~\ref{fig_dispersion_2circular} we can see that the dispersion
diagram has been modified due to the change of inclusion. The strong
coupling between two elements introduces additional guided modes to
appear in dispersion diagram. Such phenomenon has also been studied
for dielectric (non-dispersive) nanorods previously \cite{Chigrin}.
The distributions of magnetic field for selected guided modes as
marked in Fig.~\ref{fig_dispersion_2circular} are plotted in
Fig.~\ref{fig_field_2circular}.
The modes (a), (c) and (d) are even modes while (b) and (e) are odd
modes.

\begin{figure}[t]
\centering
\includegraphics[width=8.6cm]{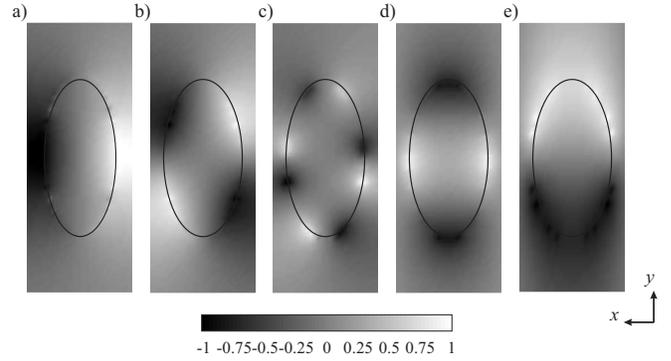}
\caption{Normalised distributions of magnetic field corresponding to
different guided modes as marked in
Fig.~\ref{fig_dispersion_elliptical}: (a), (d): even modes, and (b),
(c), (e): odd modes. The structure is infinite along $x$-direction.
(Note that the coordinate has been rotated 90 degrees anti-clockwise
from Fig.~\ref{fig_domain} for better presentation of the figure.)}
\label{fig_field_elliptical}
\end{figure}

The dispersion diagram for a single elliptical element as inclusion
is shown in Fig.~\ref{fig_dispersion_elliptical}.
It can be seen that more guided modes appear which is caused by the
change of inclusion's geometrical shape from circular to elliptical.
The frequency corresponding to the lowest mode has been lowered due
to the increase of inclusion's volume. The distributions of magnetic
field for selected guided modes are plotted in
Fig.~\ref{fig_field_elliptical}.
The modes (a) and (d) are even modes and (b), (c) and (e) are odd
modes, respectively.

\section{Wave Propagation in Plasmonic Waveguides formed by Finite Number of Elements}
In order to study wave propagations in plasmonic waveguides formed
by a finite number of silver nanorods, we have replaced PBCs in
$x$-direction with PMLs and added additional cells for the free
space region to the simulation domain. The number of nanorods under
study is seven. The spacing (pseudo-period) between adjacent
elements remains the same as for infinite structures considered in
the previous section. For a single mode excitation, we choose the
frequency of corresponding mode from dispersion diagram, and excite
sinusoidal sources at different locations with respect to the
symmetry of different guided modes at one end of the waveguides.

\begin{figure}[t]
\centering \includegraphics[width=8.6cm]{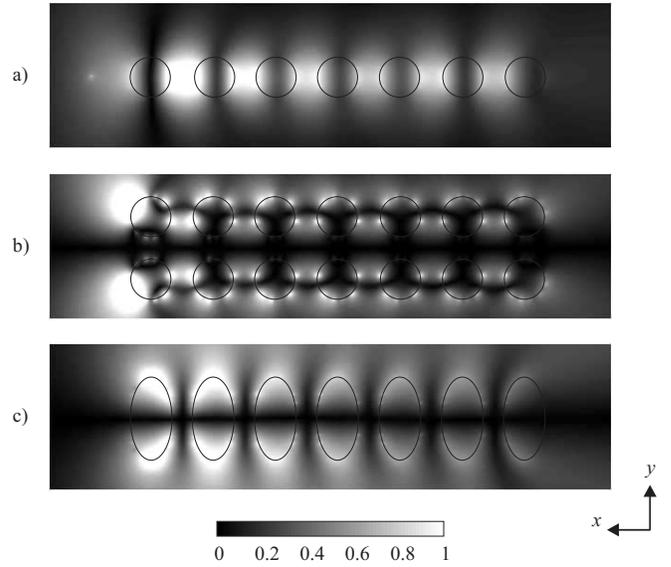}
\caption{Normalised distributions of magnetic field intensity
corresponding to different guided modes for seven-element plasmonic
waveguides formed by (a) a single row of circular nanorods (the
corresponding eigen mode is shown in
Fig.~\ref{fig_field_1circular}(a)) (b) two rows of circular nanorods
arranged in square lattice (the corresponding eigen mode is shown in
Fig.~\ref{fig_field_2circular}(c)) (c) a single row of elliptical
nanorods (the corresponding eigen mode is shown in
Fig.~\ref{fig_field_elliptical}(e)). (Note that the coordinate has
been rotated 90 degrees anti-clockwise from Fig.~\ref{fig_domain}
for better presentation of the figure.)} \label{fig_propagation}
\end{figure}

For the plasmonic waveguides formed by different types of
inclusions, we have chosen certain eigen modes: mode
Fig.~\ref{fig_field_1circular}(a) for a single row of circular
cylinders, mode Fig.~\ref{fig_field_2circular}(d) for two rows of
circular cylinders, and mode Fig.~\ref{fig_field_elliptical}(e) for
a single row of elliptical cylinders. The distributions of magnetic
field intensity for different waveguides operating in these guided
modes are plotted in Fig.~\ref{fig_propagation}.
The field plots are taken after the steady state is reached in
simulations. It is clearly seen that single guided modes are coupled
into these waveguides but the excitation of certain modes highly
depends on the symmetry of field patterns. The energy that can be
coupled into the waveguides also depends on the matching between
source and the plasmonic waveguide.

\section{Conclusions}
In conclusion, we have developed a conformal dispersive FDTD method
for the modelling of plasmonic waveguides formed by an array of
periodic infinite-long silver cylinders at optical frequencies. The
conformal scheme is based on effective permittivities and its main
advantage is that since conventional orthogonal FDTD grid is
maintained in simulations, no numerical instability is introduced.
The material frequency dispersion is taken into account using an
auxiliary differential equation method. The comparison of dispersion
diagrams for one-dimensional periodic silver cylinders calculated
using the conformal dispersive FDTD method, the conventional
dispersive FDTD method with staircase approximations and the
frequency domain embedding method demonstrates the accuracy of the
proposed method. It is shown that by adding additional element or
changing the geometry of inclusions, the corresponding dispersion
diagram can be modified. Numerical simulations of plasmonic
waveguides formed by seven elements show that the eigen modes in
infinite structures can be excited but highly depend on the symmetry
of field patterns of certain modes. Further work includes the
investigation of the effects of different number of elements in
plasmonic waveguides on guided modes, and the calculation of group
velocity of different modes propagating in these waveguides.
Although results presented in this paper have been focused at
optical frequencies, with future advances in microwave plasmonic
materials, novel applications can be found in the designs of small
antenna and efficient absorbers.

\section*{Acknowledgment}
The authors would like to thank Mr. N. Giannakis for providing
simulation results using the embedding method, and thank Dr. Pavel
Belov for helpful discussions. The authors would also like to thank
reviewers for their valuable comments and suggestions.

\balance


\begin{thebibliography}{99}
\bibitem{Joannopoulos}
J. D. Joannopoulos, R. D. Meade, and J. N. Winn, Photonic crystals:
Molding the flow of light, Princeton U. Press, Princeton, N.J.,
1995.

\bibitem{Qiu1}
G. Qiu, F. Lin, and Y. Li, ``Complete two-dimensional bandgap of
photonic crystals of a rectangular Bravais lattice,'' \textit{Opt.
Commun.}, vol. 219, pp. 285-288, 2003.

\bibitem{Fan1}
S. Fan, J. Winn, A. Devenyi, J. C. Chen, R. D. Meade, and J. D.
Joannopoulos, ``Guided and defect modes in periodic dielectric
waveguides,'' \textit{J. Opt. Soc. Am. B}, vol. 12, pp. 1267-1272,
1995.

\bibitem{Chigrin}
D. Chigrin, A. Lavrinenko, and C. Sotomayor Torres, ``Nanopillars
photonic crystal waveguides,'' \textit{Opt. Express}, vol. 12, pp.
617-622, 2004.

\bibitem{Shefer}
J. Shefer, ``Periodic cylinder arrays as transmission lines,''
\textit{IEEE Trans. Microwave Theory Tech.}, vol. 11, pp. 55-61,
Jan. 1963.

\bibitem{Munk}
B. A. Munk, D. S. Janning, J. B. Pryor, and R. J. Marhefka,
``Scattering from surface waves on finite FSS,'' \textit{IEEE Trans.
Antennas Propagat.}, vol. 49, pp. 1782-1793, Dec. 2001.

\bibitem{Mailloux}
R. J. Mailloux, ``Antenna and wave theories of infinite Yagi-Uda
arrays,'' \textit{IEEE Trans. Antennas Propagat.}, vol. 13, pp.
499-506, Jul. 1965.

\bibitem{Yaghjian}
A. D. Yaghjian, ``Scattering-matrix analysis of linear periodic
arrays,'' \textit{IEEE Trans. Antennas Propagat.}, vol. 50, pp.
1050-1064, Aug. 2002.

\bibitem{Weeber}
J. C. Weeber, A. Dereux, C. Girard, J. R. Krenn and J. P. Goudonnet,
``Plasmon polaritons of metallic nanowires for controlling submicron
propagation of light,'' \textit{Phys. Rev. B}, vol. 60, pp.
9061-9068, 1999.

\bibitem{Lamprecht}
B. Lamprecht, J. R. Krenn, G. Schider, H. Ditlbacher, M. Salerno, N.
Felidj, A. Leitner, F. R. Aussenegg, and J. C. Weeber, ``Surface
plasmon propagation in microscale metal stripes,'' \textit{Appl.
Phys. Lett.}, vol. 79, pp. 51-53, 2001.

\bibitem{Yatsui}
T. Yatsui, M. Kourogi, and M. Ohtsu , ``Plasmon waveguide for
optical far/near-field conversion,'' \textit{Appl. Phys. Lett.},
vol. 79, pp. 4583-4585, 2001.

\bibitem{Zia}
R. Zia, M. D. Selker, P. B. Catrysse, and M. L. Brongersma,
``Geometries and materials for subwavelength surface plasmon
modes,'' \textit{J. Opt. Soc. Am. A}, vol. 21, pp. 2442-2446, 2004.

\bibitem{Charbonneau}
R. Charbonneau, N. Lahoud, G. Mattiussi, and P. Berini,
``Demonstration of integrated optics elements based on long-ranging
surface plasmon polaritons,'' \textit{Opt. Express}, vol. 13, pp.
977-984, 2005.

\bibitem{Pile}
D. F. P. Pile, D. K. Gramotnev, ``Channel plasmon-polariton in a
triangular groove on a metal surface,'' \textit{Opt. Lett.}, vol.
29, pp. 1069-1071, 2004.

\bibitem{Quinten}
M. Quinten, A. Leitner, J. R. Krenn, and F. R. Aussenegg,
``Electromagnetic energy transport via linear chains of silver
nanoparticles,'' \textit{Opt. Lett.}, vol. 23, pp. 1331-1333, 1998.

\bibitem{Brongersma}
M. L. Brongersma, J. W. Hartman, and H. A. Atwater,
``Electromagnetic energy transfer and switching in nanoparticle
chain arrays below the diffraction limit,'' \textit{Phys. Rev. B},
vol. 62, pp. 16356-16359, 2000.

\bibitem{Youngs}
T. J. Shepherd, C. R. Brewitt-Taylor, P. Dimond, G. Fixter, A.
Laight, P. Lederer, P. J. Roberts, P. R. Tapster, and I. J. Youngs,
``3D microwave photonic crystals: Novel fabrication and
structures,'' Electron. Lett., vol. 34, pp. 787-789, 1998.

\bibitem{Taflove}
A. Taflove, \textit{Computational Electrodynamics: The Finite
Difference Time Domain Method}. Norwood, MA: Artech House, 1995.

\bibitem{Maier}
S. A. Maier, P. G. Kik, and H. A. Atwater, ``Optical pulse
propagation in metal nanoparticle chain waveguides,'' \textit{Phys.
Rev. B}, vol. 67, pp. 205402, 2003.

\bibitem{Saj}
W. M. Saj, ``FDTD simulations of 2D plasmon waveguide on silver
nanorods in hexagonal lattice,'' \textit{Opt. Express}, vol. 13, no.
13, pp. 4818-4827, June 2005.

\bibitem{Hwang}
K.-P. Hwang, and A. C. Cangellaris, ``Effective permittivities for
second-order accurate FDTD equations at dielectric interfaces,''
\textit{IEEE Microwave Wirel. Compon. Lett.}, vol. 11, pp. 158-160,
2001.

\bibitem{HaoConformal}
Y. Hao, and C. J. Railton, ``Analyzing electromagnetic structures
with curved boundaries on Cartesian FDTD meshes,'' \textit{IEEE
Trans. Microwave Theory Tech.}, vol. 46, pp. 82-88, Jan. 1998.

\bibitem{Luebbers1}
R. Luebbers, F. P. Hunsberger, K. Kunz, R. Standler, and M.
Schneider, ``A frequency-dependent finite-difference time-domain
formulation for dispersive materials'', \textit{IEEE Trans.
Electromagn. Compat.}, vol. 32, pp. 222-227, Aug. 1990.

\bibitem{Gandhi1}
O. P. Gandhi, B.-Q. Gao, and J.-Y. Chen, ``A frequency-dependent
finite-difference time-domain formulation for general dispersive
media,'' \textit{IEEE Trans. Microwave Theory Tech.}, vol. 41, pp.
658-664, Apr. 1993.

\bibitem{Sullivan1}
D. M. Sullivan, ``Frequency-dependent FDTD methods using Z
transforms,'' \textit{IEEE Trans. Antennas Propagat.}, vol. 40, pp.
1223-1230, Oct. 1992.

\bibitem{Kaneda}
N. Kaneda, B. Houshmand, and T. Itoh, ``FDTD analysis of dielectric
resonators with curved surfaces,'' \textit{IEEE Trans. Microwave
Theory Tech.}, vol. 45, pp. 1645-1649, Sep. 1997.

\bibitem{Lee}
J.-Y Lee and N.-H Myung, ``Locally tensor conformal FDTD method for
modeling arbitrary dielectric surfaces,'' \textit{Microw. Opt. Tech.
Lett.}, vol. 23, pp. 245-249, Nov. 1999.

\bibitem{Mohammadi}
A. Mohammadi, and M. Agio, ``Contour-path effective permittivities
for the two-dimensional finite-difference time-domain method,''
\textit{Opt. Express}, vol. 13, pp. 10367-10381, 2005.

\bibitem{Inglesfield}
J. E. Inglesfield, ``A method of embedding,'' \textit{J. Phys. C:
Solid State Phys.}, vol. 14, pp. 3795-3806, 1981.

\bibitem{Hildebrand}
F. B. Hildebrand, \textit{Introduction to Numerical Analysis}. New
York: Mc-Graw-Hill, 1956.

\bibitem{ChanPBC}
C. T. Chan, Q. L. Yu, and K. M. Ho, ``Order-N spectral method for
electromagnetic waves,'' \textit{Phys. Rev. B}, vol. 51, pp.
16635-16642, 1995.

\bibitem{Holter}
H. Holter and H. Steyskal, ``Infinite phased-array analysis using
FDTD periodic boundary conditions-pulse scanning in oblique
directions,'' \textit{IEEE Trans. Antennas Propagat.}, vol. 47, pp.
1508-1514, 1999.

\bibitem{Turner}
M. Turner and C. Christodoulou, ``FDTD analysis of phased array
antennas,'' \textit{IEEE Trans. Antennas Propagat.}, vol. 47, pp.
661-667, 1999.

\bibitem{Prescott}
D. T. Prescott and N. V. Shuley, ``Extensions to the. FDTD method
for the analysis of innitely periodic arrays,'' \textit{IEEE
Microwaves and Guided Waves Letters}, vol. 4, pp. 352-354, Oct.
1994.

\bibitem{Ren}
J. R. Ren, O. P. Gandhi, L. R. Walker, J. Fraschilla, and C. R.
Boerman, ``Floquet-based FDTD analysis of two-dimensional phased
array antennas,'' \textit{IEEE Microwave and Guided Wave Letters},
vol. 4, pp. 109-111, 1994.

\bibitem{Roden}
J. A. Roden, S. D. Gedney, M. P. Kesler, J. G. Maloney, and P. H.
Harms, ``Time-domain analysis of periodic structures at oblique
incidence: Orthogonal and nonorthogonal FDTD implementations,''
\textit{IEEE trans. Microwave Theory and Techniques}, vol. 46, pp.
420-427, 1998.

\bibitem{Fan2}
S. Fan, P. R. Villeneuve and J. D. Joannopoulos, ``Large
omnidirectional band gaps in metallodielectric photonic crystals,''
\textit{Phys. Rev. B}, vol. 54, pp. 11245-11251, 1996.

\bibitem{Qiu2}
M. Qiu and S. He, ``A nonorthogonal finite-difference time-domain
method for computing the band structure of a two-dimensional
photonic crystal with dielectric and metallic inclusions,''
\textit{J. Appl. Phys.}, vol. 87, pp. 8268-8275, 2000.

\bibitem{Berenger}
J. R. Berenger, ``A perfectly matched layer for the absorption of
electromagnetic waves,'' \textit{J. Computat. Phys.}, vol. 114, pp.
185-200, Oct. 1994.

\bibitem{Johnson}
P. B. Johnson and R. W. Christy, ``Optical constants of the noble
metals,'' \textit{Phys. Rev. B}, vol. 6, pp. 4370-4379, 1972.

\bibitem{Giannakis}
N. Giannakis, J. Inglesfield, P. Belov, Y. Zhao, and Y. Hao,
``Dispersion properties of subwavelength waveguide formed by silver
nanorods,'' Photon06, September 3-7, 2006, Manchester, UK.

\bibitem{Song}
W. Song, Y. Hao, and C. Parini, ``Comparison of nonorthogonal and
Yee's FDTD schemes in modelling photonic crystals,'' submitted to
\textit{Opt. Express}, 2006.

\end{thebibliography}
\end{document}